\documentstyle[twocolumn,prl,aps,epsf]{revtex}
\begin{document}
\draft
\title{Nonlinear Competition Between Small and Large Hexagonal Patterns}
\author{Mary Silber} 
\address{Department of Engineering Sciences and Applied Mathematics,
Northwestern University, Evanston, IL 60208 USA}
\author{Michael R.E. Proctor} 
\address{Department of Applied Mathematics and Theoretical Physics,
University of Cambridge, Cambridge CB3 9EW UK}
\maketitle
\begin{abstract}
Recent experiments by Kudrolli, Pier and Gollub \cite{kg} on surface
waves, parametrically excited by two-frequency forcing, show a
transition from a small hexagonal standing wave pattern to a
triangular ``superlattice'' pattern. We show that generically the
hexagons and the superlattice wave patterns bifurcate {\it
simultaneously} from the flat surface state as the forcing amplitude
is increased, and that the experimentally-observed transition can be
described by considering a low-dimensional bifurcation problem. A
number of predictions come out of this general analysis.
\end{abstract}
\newcommand{\be}{\begin{equation}\label}
\newcommand{\ee}{\end{equation}}
\newcommand{\bea}{\begin{eqnarray}\label}
\newcommand{\eea}{\end{eqnarray}}

\begin{figure}
\centerline{
\epsfxsize=115pt
\epsffile{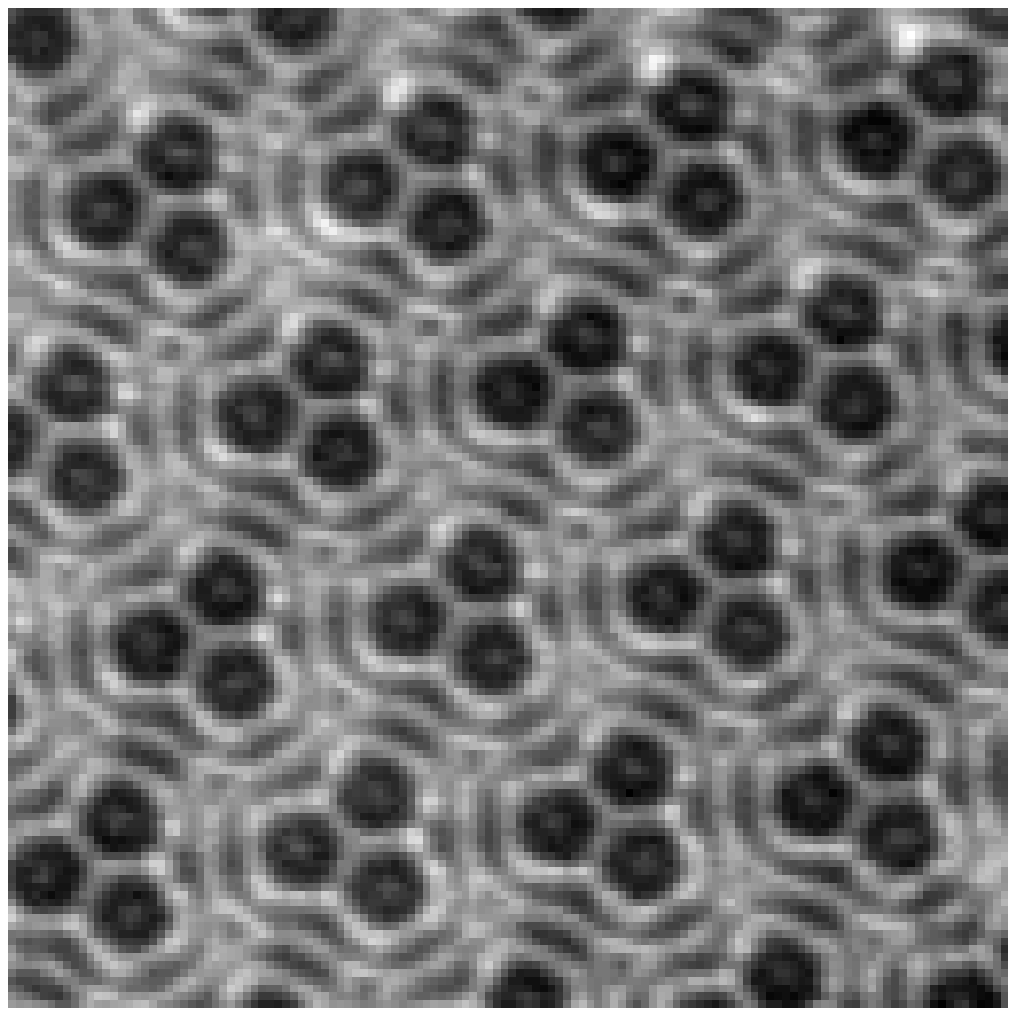}
\hskip 0.1truein
\epsfxsize=115pt
\epsffile{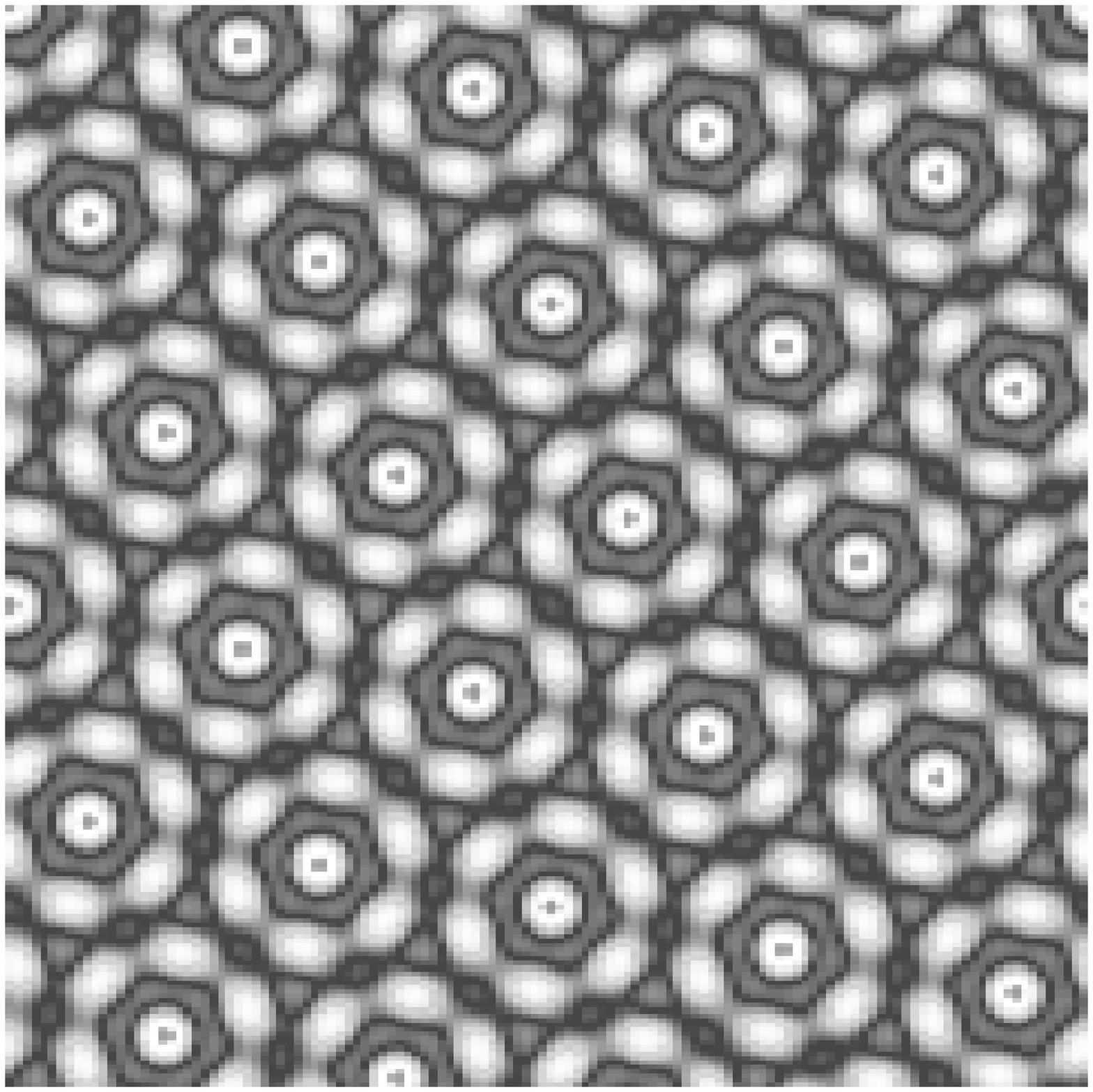}} 
\vskip .1truein
\caption{On the left is a blow up of the experimental superlattice
standing wave pattern \protect\cite{kg} (courtesy of Kudrolli, Pier
and Gollub). We show a region with area $\sim 1/30$ of the circular
container's cross-sectional area.  Note that the pattern is periodic
on a hexagonal lattice, and that it has triangular symmetry. On the
right is a hexagonal superlattice Turing pattern obtained from a
numerical integration of a reaction-diffusion system
\protect\cite{judd} (courtesy of Judd).}
\label{figa}
\end{figure}

In recent years there has been considerable interest in pattern
formation in spatially extended nonequilibrium systems
\cite{cross}. These studies have identified universal mechanisms for
the formation of certain patterns. For instance, the origin of the
ubiquitous hexagonal pattern can be traced to a symmetry-breaking
instability of a spatially homogeneous state that occurs in the
presence of certain Euclidean symmetries, which play a central role in
determining the possible nonlinear evolution of the instability. In
this letter we use similar ideas to investigate a family of hexagonal
and triangular patterns which are born in the same symmetry-breaking
instability that gives rise to simple hexagons, but have structure on
two disparate length scales \cite{dionne,dss}.  Kudrolli {\it et al.}
\cite{kg} call such structures ``superlattice patterns'';
Fig.~\ref{figa} gives two examples. Current interest in superlattice
patterns is sparked by recent observations of their formation in
experiments on parametrically excited surface waves
\cite{kg} and in nonlinear optics experiments 
\cite{pampaloni}, as well as in a
study of Turing patterns in reaction-diffusion systems
\cite{judd}.

In this letter we show that superlattice patterns can arise {\it
directly} from the spatially homogeneous state via a transcritical
bifurcation.  Moreover, this occurs in generic pattern-forming systems
for which there is only {\it one} unstable wavenumber. We show that
the triangular superlattice pattern differs from simple patterns such
as hexagons and stripes, and complex quasi-patterns, because it is
characterized by {\it both} an amplitude and a {\it phase} which
depend on the distance from the bifurcation point.  In order to
investigate the phase associated with this state we find that it is
necessary to include high order resonant interaction terms in the
amplitude equations. When these terms are neglected the problem
becomes degenerate and superlattice patterns such as those in
Fig.~\ref{figa} become just two isolated examples in a continuum of
states that have varying degrees of symmetry. Thus high order terms
are essential to understanding the formation of superlattice patterns.

While much of our analysis is quite general, our presentation will
focus on the recent experiments of
\cite{kg}, in which surface waves are parametrically
excited by subjecting a fluid layer to a periodic vertical
acceleration that involves two rationally-related frequencies. In the
classic Faraday problem, which has single frequency forcing, the first
modes to lose stability with increased acceleration are subharmonic
with respect to the vibration frequency
\cite{miles}. With two frequencies, harmonic response can
occur\cite{ef,bet}.  In particular, the triangular superlattice
pattern in Fig.~\ref{figa} was obtained with an acceleration of the
form $f(t)= a(\cos(\chi)\cos(6\omega t)+
\sin(\chi)\cos(7\omega t+\phi)). $ 
In the experiments, a hexagonal standing wave pattern is produced at
the onset of instability, then, as the acceleration is increased,
there is a transition to the superlattice pattern depicted in
Fig.~\ref{figa}
\cite{kg}. Comparison of the spatial
Fourier transform of the onset hexagons and the superlattice patterns
reveals that the new state is formed from the nonlinear interaction of
twelve prominent Fourier modes whose wavenumbers have the same modulus
$k_c$ and lie on a hexagonal lattice with fundamental wavenumber
$k_c/\sqrt{7}$.  In this letter we address the transition between
hexagons and superlattice patterns by considering a degenerate
bifurcation problem akin to that which successfully explains the
transition between hexagons and rolls in a variety of systems.  A
number of concrete predictions about the form, origin and stability
properties of the superlattice patterns come out of this general
symmetry-based analysis.

\begin{figure}
\centerline{
\epsfxsize=150pt
\epsffile{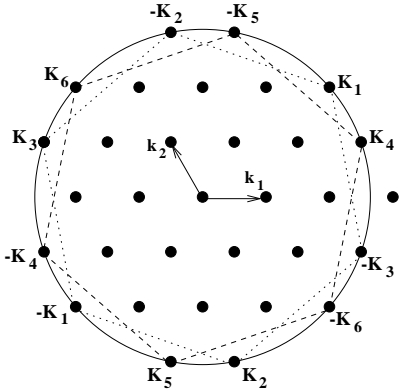} 
}
\caption{Intersection of the critical circle $|{\bf k}|=k_c$ with the
${\bf k}-$space hexagonal lattice generated by ${\bf k}_1$, ${\bf
k}_2$. In this example $\alpha=3,\ \beta=2$ in (\protect\ref{eq:ck}).
The wave vectors $\pm {\bf K}_1,\ldots,\pm {\bf K}_3$ lie at
the vertices of a hexagon, as do  $\pm {\bf
K}_4,\ldots,\pm {\bf K}_6$.}
\label{fig1}
\end{figure}

The experimentally observed patterns are put into a theoretical
framework by restricting to solutions that tile the plane
in a hexagonal fashion.
The time-periodic forcing in the Faraday problem leads us to a formulation
in terms of a stroboscopic map. Since we seek spatially
periodic solutions, we express all fields in terms of double Fourier
series; for example, the height of the free surface, at time $t=mT$,
is \be{eq:df} \zeta_m({\bf x})={\rm Re}\Bigl(\sum_{{\bf n}\in{\bf
Z}^2}w_{\bf n}(mT)\ e^{i(n_1{\bf k}_1+n_2{\bf k}_2)\cdot{\bf
x}}\Bigr)\ , \ee where $T$ is the period of the forcing. We obtain
patterns periodic on a hexagonal lattice when the primitive vectors
${\bf k}_1,{\bf k}_2$ satisfy $|{\bf k}_1|=|{\bf k}_2|=k$ and ${\bf
k}_1\cdot {\bf k}_2=-{1\over 2}k^2$. A standard {\it ansatz} in
pattern selection problems is to set $k=k_c$, where $k_c$ is the
critical wavenumber of the instability. However, here we want to
investigate competition between small and large hexagonal patterns,
which we do following the approach in
\cite{dss}. Specifically, we choose $k$ so that there are twelve wave
vectors ${\bf K}_{\bf n}=n_1{\bf k_1}+n_2{\bf k_2}$ in (\ref{eq:df})
that satisfy $|{\bf K}_{\bf n}|=k_c$, {\it i.e.}, so that there exists
a co-prime integer pair ${\bf n}=(\alpha,\beta)$, $\alpha > \beta >
0$, such that \be{eq:ck} |\alpha {\bf k_1}+ \beta {\bf
k_2}|=k\sqrt{\alpha^2+\beta^2-\alpha\beta}=k_c\ .  \ee Fig.~\ref{fig1}
presents an example associated with $\alpha=3,\beta=2$,
$k_c/k=\sqrt{7}$. Note that the lattice points on the critical circle
lie at the vertices of two hexagons rotated by an angle $\theta$
relative to each other.  The angle, determined by the dot product of
${\bf K}_1=\alpha {\bf k}_1+\beta {\bf k}_2$ and ${\bf K}_4=\alpha
{\bf k}_1+(\alpha-\beta) {\bf k}_2$, satisfies
$\cos(\theta)={\alpha^2+2\alpha\beta-2\beta^2\over
2(\alpha^2-\alpha\beta+\beta^2)}$.  If only those modes associated
with one of the hexagons are excited, then the periodicity of the
pattern is dictated by $k_c$ and one recovers the standard
six-dimensional hexagonal lattice bifurcation problem. However, if all
twelve modes on the critical circle are excited, then the period of
the pattern is greater by a factor of $k_c/k$.  In order to simplify
our presentation we now set $\alpha=3,\beta=2$; the results for
general $\alpha,\beta$ are similar.

In formulating the bifurcation problem, we assume that the flat fluid
surface loses stability to harmonic waves of wavenumber $k_c$ as a
bifurcation parameter $\lambda$ is increased through zero. Thus the
Faraday instability sets in when a Floquet multiplier $\mu$ crosses
the unit circle at $\mu=1$. In our formulation there are twelve (real)
Fourier modes in (\ref{eq:df}) that are neutrally stable at
$\lambda=0$; all others are damped. In this case the stroboscopic map
can be reduced, near the onset of instability, to a
twelve-dimensional center manifold spanned by the critical Fourier
modes.  Let $z_j(m)$, $j=1,\dots,6$, be the complex amplitude, at
$t=mT$, of the Fourier mode $e^{i{\bf K}_j\cdot{\bf x}}$, with
$\overline{z}_j$ being the amplitude of $e^{-i{\bf K}_j\cdot{\bf x}}$.
Here the ${\bf K}_j$ are labeled as in Fig.~\ref{fig1}.

The possible nonlinear terms in the stroboscopic map 
${\bf z}(m+1)={\bf f}({\bf z}(m))$,
${\bf z}\equiv(z_1,\ldots,z_6)$,  
are restricted by
the symmetries of the problem.
The first component of ${\bf f}$ has the general form
\bea{eq:f1} f_1&=&h_1({\bf u},{\bf q},\overline{\bf q})z_1+ h_2({\bf
u},{\bf q},\overline{\bf q})\overline{z}_2\overline{z}_3\nonumber\\ &&
+ c_1 z_3^2 z_4^2
\overline{z}_6 +c_2
\overline{z}_1z_2
z_4\overline{z}_5^2 +{\cal O}(|{\bf z}|^{6}),
\eea where ${\bf u}\equiv
(|z_1|^2,|z_2|^2,|z_3|^2,|z_4|^2,|z_5|^2,|z_6|^2)$, and ${\bf q}\equiv
(z_1z_2z_3,z_4z_5z_6)$. The discrete hexagonal symmetries place
further restrictions on the functions $h_1, h_2$ and also determine
the other components of ${\bf f}$ from $f_1$ \cite{dss}. At this point
we note that if the resonant interaction terms, with coefficients
$c_1$ and $c_2$, are neglected, then the phases $\phi_j$ associated
with each $z_j\equiv R_je^{i\phi_j}$ only enter the problem 
through the values of the total phases
$\Phi_1\equiv \phi_1+\phi_2+\phi_3$ and
$\Phi_2\equiv\phi_4+\phi_5+\phi_6$. As a
result the cubic truncation of~(\ref{eq:f1}) admits solutions in the
form of two hexagons rotated relative to each other by $\theta\approx
22^\circ$ (for $\alpha=3,\beta=2$), and translated relative to each
other by {\it arbitrary} amounts.
Kudrolli, {\it et al.}
\cite{kg}  noted that only a restricted set of relative translations 
of the two hexagons leads to a superlattice pattern which retains
triangular symmetry. Their phenomenological description did not,
however, give a mechanism for the selection of a particular shift.  We
now investigate an alternative description of the triangular
superlattice pattern which has the advantage that is in fact
consistent with the inclusion of the high order resonant interaction
terms.

We focus on patterns that have at least three-fold rotational symmetry
by restricting the bifurcation problem to the subspace ${\bf
z}(m)=(u_m,u_m,u_m,v_m,v_m,v_m)$, where $u_m$ and $v_m$ are complex
Fourier amplitudes measured at time $t=mT$. First we recall some
results about period-one simple hexagons which satisfy
$(u_m,v_m)=(Re^{i\varphi},0)$ or $(0,Re^{i\varphi})$.  The amplitude
$R$ and phase $\varphi$ obey \bea{eq:simplehex} 0&=&\lambda+\epsilon
R\cos(3\varphi)+AR^2+\cdots\nonumber\\
0&=&\sin(3\varphi)\Bigl(-\epsilon+BR^2+\cdots\Bigr), \eea where
$\epsilon$, $A$ and $B$ are nonlinear coefficients that arise from the
Taylor expansions of $h_1$ and $h_2$ in (\ref{eq:f1}).  The solutions
of (\ref{eq:simplehex}) that bifurcate from the origin at $\lambda=0$
satisfy $\sin(3\varphi)=0$; the hexagons with $\varphi=0,\pm
{2\pi\over 3}$ are related by translations, as are the hexagons with
$\varphi=\pi,\pm{\pi\over 3}$. These two sets of hexagons,
`up-hexagons' ($H^+$) and `down-hexagons' ($H^-$), form the two
branches associated with a transcritical bifurcation.  In
non-Boussinesq convection, these states correspond to ones in which
fluid is rising or falling at the centers of convection cells
\cite{busse}.

\begin{figure}
\centerline{
\epsfxsize=250pt
\epsffile{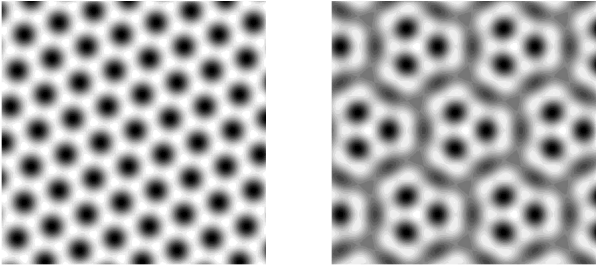} 
}
\caption{Examples of patterns, periodic on a hexagonal lattice, that
bifurcate transcritically from the flat state at $\lambda=0$. These
are plots of appropriate superpositions of critical Fourier modes with
wave vectors $\pm {\bf K}_1,\dots,\pm {\bf K}_6$ of
Fig.~\protect\ref{fig1}. The plots are, left to right, down-hexagons
($H^-$) and superlattice down-triangles ($T^-$) ({\it
cf}. Fig.\protect\ref{figa}). The critical wavenumber $k_c$ dictates
the size of the small scale structure evident in the superlattice
pattern.}
\label{fig2}
\end{figure}

Next we consider patterns that involve all six complex Fourier modes
$z_1,\dots,z_6$, with $z_j=\rho e^{i\psi}$.  We find that the
amplitude $\rho$ and phase $\psi$ satisfy equations of the form, {\it
cf.} (\ref{eq:simplehex}),
\bea{eq:superhex} 0&=&\lambda +\epsilon \rho\cos(3\psi)+\widetilde{A}
\rho^2+\cdots\\
0&=&\sin(3\psi)\Bigl(-\epsilon+\widetilde{B}\rho^2+\cdots\Bigr)
+(c_1-c_2)\rho^3\sin(2\psi)+\cdots\ \nonumber \eea 
In this case there
are two distinct types of solutions that bifurcate from the origin at
$\lambda=0$; those with phase $\psi=0,\pi$ and those with $\psi\approx
\pm\pi/3,\pm 2\pi/3$. Note that
the phase $\psi$ associated with the latter solutions depends on the
amplitude $\rho$, which in turn depends on the distance from the
bifurcation point.  The solutions with $\psi=0,\pi$ have hexagonal
symmetry; in
\cite{dss} they are referred to as ``super hexagons''. As
with simple hexagons,  super hexagons bifurcate transcritically
with the two parts of the branch satisfying $\psi=0$ and $\psi=\pi$,
respectively. The solutions satisfying $\psi\approx
\pm\pi/3,\pm 2\pi/3$ have only triangular symmetry, and are new. 
The triangular solutions with $\psi\approx 2\pi/3$ and $\psi\approx
-\pi/3$ form the two parts of a transcritical branch; a rotation of
these patterns by $\pi$ changes the sign of $\psi$. These solutions
have structure identical to the superlattice harmonic state
described in
\cite{kg}: compare the triangular superlattice pattern in
Fig.~\ref{figa} with that in Fig.~\ref{fig2}. Moreover, we know that
the solutions with $\cos(3\varphi)>0$ in (\ref{eq:simplehex}) and
$\cos(3\psi)>0$ in (\ref{eq:superhex}) bifurcate in the same direction
from the origin, as do the branches with
$\cos(3\varphi),\cos(3\psi)<0$.

Finally, we address the experimentally observed transition between the
patterns in Fig.~\ref{fig2}. Since all of the known primary solution
branches are unstable at bifurcation, due to the presence of the
quadratic nonlinear term in (\ref{eq:f1}), we assume that the
coefficient $\epsilon$ of the quadratic term satisfies $|\epsilon|\ll
1$. This allows us to investigate stable small amplitude solutions,
and the transitions between them. If we truncate~(\ref{eq:f1}) at
cubic order, thereby neglecting the high-order resonant terms with
coefficients $c_1,$ $c_2$, then the superlattice patterns are at best
neutrally stable.  Specifically, while we expect a (multiplicity two)
unit multiplier associated with translations of the pattern, the extra
multiplier (also of multiplicity two) results from a symmetry of the
{\it truncated} equations that allows a {\it relative} translation of
the two rotated hexagons that make up the pattern. As noted above,
this symmetry is broken when the resonant terms are included; the
Floquet multiplier $\mu$ then moves off the unit circle. 
In
particular, $\mu>1$ for the super hexagons if $4c_1+5c_2<0$
\cite{dss}. Analogously, we can show that $\mu<1$ for the triangular
superlattice pattern if $4c_1+5c_2<0$.  Thus if the superlattice
patterns are {\it neutrally} stable when the high-order resonant terms
are neglected, then we expect one of them is stable and the other is
in fact unstable. This is consistent with the experimental
observations of Kudrolli, {\it et al.} \cite{kg} who only observe the
triangular superlattice pattern.

\begin{figure}
\centerline{
\epsfxsize=200pt
\epsffile{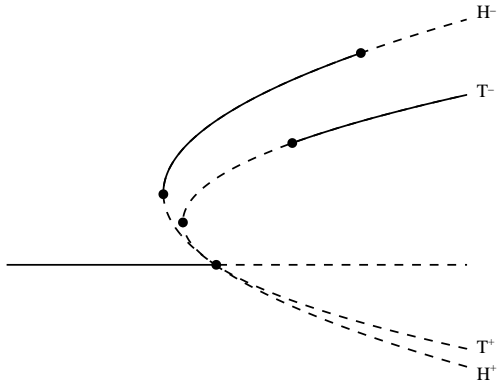} 
}
\caption{Part of the bifurcation diagram associated with the
bifurcation problem (\protect\ref{eq:ex}) for $-1\ll \epsilon <
0$. Bifurcation points are indicated by solid circles. Only branches
of simple hexagons $(H^\pm)$ and superlattice triangles $(T^\pm$) are
indicated; solid lines correspond to stable solutions. Branches
produced in secondary bifurcations, and {\it unstable} primary
branches are not shown.}
\label{fig3}
\end{figure}

In Fig.~\ref{fig3}, we present part of a bifurcation diagram
associated with the stroboscopic map. Specifically, we keep all
terms through cubic and the essential quintic terms: 
\bea{eq:ex} f_1&=&(1+\lambda)z_1+\epsilon
\overline{z}_2\overline{z}_3+ (a_1|z_1|^2+a_2|z_2|^2+a_2|z_3|^2)z_1\nonumber\\
&&+(a_4|z_4|^2+a_5|z_5|^2+a_6|z_6|^2)z_1 \\
&&+c_1z_3^2z_4^2\overline{z}_6+c_2\overline{z}_1 z_2z_4\overline{z}_5^2\
,\nonumber \eea 
and assume that the cubic coefficients satisfy the following two 
inequalities:
\bea{eq:ineq} 
a_1+2a_2&<& -|a_4+a_5+a_6|\\ a_2-a_1&>& \sqrt{{
(a_4-a_5)^2+(a_4-a_6)^2+(a_5-a_6)^2\over 2}}\ ,
\nonumber \eea 
with $4c_1+5c_2<0$. Conditions (\ref{eq:ineq}) ensure that
superlattice patterns eventually supersede the simple hexagons, and
that the stripe pattern is never stabilized.

This example demonstrates that it is possible to reproduce the type of
transition reported in \cite{kg} within the framework of a generic
bifurcation problem on a hexagonal lattice. Moreover, a number of
predictions come out of this general analysis. Specifically, we find
that {\it both} the transition between the the flat state and the
small hexagons, and the transition between the hexagons and the
superlattice pattern are hysteretic, with the region of bistability
between the hexagons and the superlattice pattern exceeding that
between the hexagons and the flat state. While there is no clear
detection of hysteresis in the experiments, the transition from
hexagons to the superlattice pattern is reported to occur via domains
of the superlattice pattern growing, with increased acceleration,
until they fill the container \cite{kg}. Coexisting domains of
different patterns, possibly with slowly moving fronts between them,
is a common manifestation of bistability of patterns in extended
systems. Another prediction that comes out of our general analysis is
that for $\epsilon<0$ the transition is between the down-states
depicted in Fig.~\ref{fig2}, while if $\epsilon>0$ the transition is
between up-states. While in Rayleigh-B\'enard convection there is a
clear distinction between up- and down-states, this distinction may be
a subtle one in the Faraday experiment since standard imaging
techniques rely on reflection of light from the surface of the fluid,
so that surface peaks are not distinguishable from troughs. The
nonlinearity inherent in the imaging may also present challenges to
extracting the amplitude-dependent phase that we predict is associated
with the triangular superlattice patterns.

While the foregoing analysis shows how it is possible to achieve the
experimentally observed transitions within a general framework, many
of the specific predictions of the analysis still need to be
tested. Furthermore, it would be very interesting to calculate the
nonlinear coefficients in (\ref{eq:f1}) from the hydrodynamic
equations to determine whether the minimal inequalities
(\ref{eq:ineq}) are satisfied for the experimental parameters. Such
computations are quite involved in the case of two-frequency forcing
of viscous fluids. However, much progress has been made in the case of
very low viscosity fluids by Zhang and Vi\~nals \cite{vinal}, who
employ a quasi-potential approximation to the Navier-Stokes equations
to simplify the calculations.

Finally, we note that all our calculations have assumed that the
solutions are strictly periodic on a hexagonal lattice.  This is
certainly an appropriate model for observations of superlattice
patterns, but as noted in \cite{kg} quasi-patterns are observed for
other experimental parameters. A mechanism for favoring
quasi-patterns by a resonant interaction between two bifurcating
states with different horizontal wavenumbers was proposed by Edwards
and Fauve
\cite{ef}, and investigated for a Swift-Hohenberg type of
model by Lifshitz and Petrich \cite{lifshitz}. Whether a similar
resonance mechanism is responsible for the spatially-periodic
superlattice patterns is an intriguing, open question.

We thank J.P.~Gollub, A.~Kudrolli and B. Pier for discussing their
experiments with us, and S.L.~Judd, H.~Riecke, A.M.~Rucklidge and
A.C.~Skeldon for helpful discussions.  This research was supported by
a NATO collaborative research grant CRG-950227. The research of MS is
supported by an NSF CAREER award DMS-9502266. The research of MREP is
supported by the UK PPARC and EPSRC.


\begin{thebibliography}{99}
\bibitem{kg} A.~Kudrolli, B.~Pier and J.P.~Gollub, {\it Physica D}, in
press (1998).
\bibitem{cross} For a review of pattern formation, 
see M.C.~Cross and P.C.~Hohenberg, Rev. Mod. Phys. {\bf 65}, 851 (1993).
\bibitem{dionne} B.~Dionne and M.~Golubitsky, 
ZAMP {\bf 43}, 36 (1992).
\bibitem{dss} B.~Dionne, M.~Silber and A.C.~Skeldon, Nonlinearity
{\bf 10},  321 (1997). 
\bibitem{pampaloni} E.~Pampaloni, S.~Residori, S.~Soria, and
F.T.~Arecchi, Phys. Rev. Lett. {\bf 78}, 1042 (1997).
\bibitem{judd} S.L.~Judd and M.~Silber, preprint (1998)\hfil\\ 
{\tt http://xxx.lanl.gov/abs/patt-sol/9807002}.
\bibitem{miles} For a review of the Faraday problem, 
see J.W.~Miles and D.~Henderson, Annu. Rev. Fluid Mech.
{\bf 22}, 143 (1990).
\bibitem{ef} W.S.~Edwards and S. Fauve, J. Fluid Mech. {\bf 278}, 123
(1994).
\bibitem{bet} T.~Besson, W.~Stuart Edwards and L.S.~Tuckerman,
Phys. Rev. E {\bf 54}, 507 (1995).
\bibitem{busse} F.H.~Busse, Rep. Prog. Phys. {\bf 41} 1929 (1978).
\bibitem{vinal} W.~Zhang and J.~Vi\~nals, J. Fluid Mech. {\bf 341}, 225 (1997).
\bibitem{lifshitz} R.~Lifshitz and D.M.~Petrich, {\it Phys. Rev. Lett}
{\bf 79}, 1261 (1997).
\end{thebibliography}
\end{document}